\documentclass[twocolumn,english,aps,reprint,groupedaddress,notitlepage,nobibnotes,nofootinbib,preprintnumbers,showpacs]{revtex4-1}
\usepackage{graphicx,epsf,color,amsmath,multirow}
\usepackage[caption=false]{subfig}
\usepackage[yyyymmdd,hhmmss]{datetime}
\newif\ifdraft
\drafttrue
\newif\ifpreprint
\preprinttrue

\def\spa#1.#2{\left\langle#1\,#2\right\rangle}
\def\spb#1.#2{\left[#1\,#2\right]}

\def\eqn#1{Eq.~(\ref{#1})}
\def\Eqn#1{Equation~(\ref{#1})}
\def\eqns#1#2{Eqs.~(\ref{#1}) and~(\ref{#2})}

\def\NeqFour{{{\cal N}=4}}

\newcommand{\deltaQ}{\delta^{(8)} (Q)}

\def\nn{\nonumber}
\def\Section#1{\vskip .2 cm 
\noindent{\em #1:}}

\begin{document}

\preprint{
\centerline{\hskip -9 cm UCLA/17/TEP/107 \hfill NSF-ITP-17-266 \hskip .3cm}
}

\title{Cancelling the $U(1)$ Anomaly in the S-matrix of ${{\cal N}=4}$ Supergravity}

\author{Zvi~Bern${}^{a }$, Julio Parra-Martinez${}^{a }$ and Radu Roiban${}^{b }$}
\affiliation{
${}^a$Mani L. Bhaumik Institute for Theoretical Physics,\\
UCLA Department of Physics and Astronomy,\\
Los Angeles, CA 90095, USA\\
${}^b$Institute for Gravitation and the Cosmos,\\
Pennsylvania State University,\\
University Park, PA 16802, USA
}

\begin{abstract}
  ${{\cal N}=4}$ supergravity is understood to contain a $U(1)$ anomaly
  which manifests itself via the nonvanishing of loop-level scattering
  amplitudes that violate a tree-level charge conservation rule.  In
  this letter we provide detailed evidence that at one loop such
  anomalous amplitudes can be set to zero by the addition of a finite
  local counterterm. 
  We show
  that the same counterterm also cancels evanescent contributions
  which play an important role in the analysis of ultraviolet
  divergences in dimensionally regularized gravity.  These
  cancellations call for a reanalysis of the four-loop ultraviolet
  divergences previously found in this theory without the addition of
  such counterterms.
\end{abstract}


\maketitle

\Section{Introduction} 
Whenever a classical symmetry of a quantum field theory is broken by quantum
corrections it is said to exhibit an anomaly.  This may be either an off-shell
symmetry (i.e. of the Lagrangian) or an on-shell symmetry (i.e. of the
equations of motion or the S-matrix). Anomalies can have important physical
consequences. For example, they may render field theories nonrenormalizable and
nonunitary or they may yield nonzero loop-level S-matrix elements whose
tree-level counterparts vanish identically due to the classical symmetry. In
general, quantum theories are ambiguous up to the addition of local
counterterms and sometimes these can be chosen to remove an anomaly, perhaps at
the cost of breaking another symmetry.

In the context of extended four-dimensional supergravity theories
an interesting example of a symmetry susceptible to anomalies is given
by their duality symmetries~\cite{Cremmer1979up, N4superGrav,
  duality_dim_red}. These symmetries involve electric/magnetic duality
transformations of abelian vector fields which leave invariant the
equations of motion and, as such, are
symmetries of the on-shell type. In addition they act nonlinearly on
scalars parametrizing a $G/H$ sigma model. 
The presence of anomalies in such sigma models when coupled to fermions was
first addressed in Ref.~\cite{Moore1984dc} and revisited in the setting of
extended supergravities in Ref.~\cite{dfg}.  While the duality groups of ${\cal
N} > 4$ supergravities are expected to be preserved at the quantum level, it is
argued in Ref.~\cite{MarcusAnomaly} that the classical
$SU(1,1)$ duality group of ${\cal N} = 4$ supergravity is anomalous.  
This anomaly can be pushed into a $U(1)$ subgroup that
``sources'' certain classes of amplitudes that vanish at
tree-level~\cite{CarrascoAnomaly}.

The duality symmetries have received renewed interest due to their
implications on the ultraviolet properties of these
theories~\cite{UVDuality, ElvangKiermaier, BHSAnomalies, CarrascoAnomaly, RenataAnomalies}. The
precise connection between the anomalies in these symmetries and
ultraviolet divergences remains to be fully
unraveled~\cite{ThreeLoopN4Sugra,BHSAnomalies,RenataAnomalies}, but there are hints
that in $\NeqFour$ supergravity the two are
tied~\cite{CarrascoAnomaly, FourLoopN4Sugra, RenataConjecture}.
For instance, the amplitudes sourced by the anomaly, when inserted into
unitarity cuts at higher loops, lead to potential ultra-violet divergent
contributions~\cite{CarrascoAnomaly}, which cancel at three
loops~\cite{ThreeLoopN4Sugra}, but not at four loops~\cite{FourLoopN4Sugra}.
Another connection is the recent observation that the anomalous amplitudes are
intertwined with evanescent contributions~\cite{CurvatureSquareN4}, similar to
those which played an important role in the interpretation of the two-loop
ultraviolet divergence of pure Einstein gravity in dimensional
regularization~\cite{GoroffSagnotti,vandeVen}. Evanescent operators---such as
the Gauss--Bonnet $R^2$ operator---are those whose tree-level matrix elements
vanish identically in four but not general dimensions. Because of this
property, they can contribute to divergences in dimensional regularization
while otherwise having no physical consequences in the amplitudes, since their
effects can be removed by the addition of local
counterterms~\cite{GBPureGravity}.
Similar curvature-squared evanescent contributions also appear at one
loop in $\NeqFour$ supergravity~\cite{CurvatureSquareN4}, although
with a ultraviolet-finite coefficient.  From the perspective of the
double-copy construction of $\NeqFour$
supergravity~\cite{BCJLoop,N4Oneloop}, in terms of $\NeqFour$
super-Yang--Mills and pure Yang--Mills theory, both these and the
anomalous contributions originate in the same rational pieces
of the corresponding Yang--Mills amplitudes

Given the relation between the anomalous terms and the evanescent
contributions, and the understanding that the effects of the latter
can be absorbed into finite local counterterms, we are immediately led
to the following questions: Can we absorb the effects of the anomaly
in the one-loop scattering amplitudes into a local counterterm? More
generally, can the anomaly be cancelled in such a way?  Motivated by
these questions we explicitly computed a variety of anomalous
amplitudes in $\NeqFour$ supergravity, including infinite classes, and
show that they can indeed be set to zero by the addition of a finite
local counterterm given by the supersymmetrization of a
curvature-squared operator multiplied by scalars.  The restoration of
the anomalous $U(1)$ symmetry via the addition of a local counterterm has
already been discussed in Ref.~\cite{BHSAnomalies} in the context of
the ungauge-fixed effective action.  Here we address a related, but
somewhat different, issue of finding a specific counterterm that
removes anomalous scattering amplitudes.

\Section{Review}
The physical complex scalar of ${\cal N}=4$ supergravity
parametrizes the coset $SU(1,1)/U(1)$; the choice of parameterization
determines the anomalous $U(1)$. In a classically-equivalent formulation with a
global $SU(1,1)$ symmetry and an auxiliary $U(1)$ gauge
symmetry~\cite{Cremmer1979up} it is the latter symmetry which is anomalous.
The former framework is recovered upon gauge fixing~\cite{dfg, gdw}. 
In the so-called $SU(4)$ gauge fixing the $SU(1,1)$ transformations are: (a)
shifting the physical scalar $\tau$, (b) rescaling $\tau$ while rescaling oppositely
the vector fields and (c) nonlinearly transforming $\tau$ while chirally-rotating
the fermions and dualizing the vector fields.
It is the third generator which is anomalous.

On shell, the spectrum of $\NeqFour$ supergravity consists of two different
supermultiplets which can be represented using an on-shell
superspace~\cite{NairSuperspace} as
\begin{align}
\Phi^+ & =  h^{++} + \eta^A \psi^+_A + \frac{1}{2!} \eta^A\eta^B A^+_{AB} \nonumber\\
& \null + \frac{1}{3!} \eta^A\eta^B\eta^C \varepsilon_{ABCD}\, \chi^{+D} 
+ \frac{1}{4!}\eta^A\eta^B\eta^C\eta^D \varepsilon_{ABCD}\, \bar{t}\,, \nonumber\\
\Phi^- & =  t + \eta^A \chi^-_A + \frac{1}{2!} \eta^A\eta^B A^-_{AB}  
+ \frac{1}{3!} \eta^A\eta^B\eta^C \varepsilon_{ABCD}\, \psi^{-D} \nonumber \\
& \null + \frac{1}{4!}\eta^A\eta^B\eta^C\eta^D \varepsilon_{ABCD}\, h^{--}\,.
\label{Multiplets}
\end{align}
The indices $A,B,C,D$ are $SU(4)$ $R$-symmetry indices.  The $\Phi^+$ multiplet
contains the positive-helicity graviton $h^{++}$ , the four positive-helicity
gravitinos $\psi^+_A$, and so forth down to the complex scalar $\bar t$. The
second supermultiplet is the CPT conjugate to first one and contains the
negative-helicity graviton $h^{--}$ and the conjugate scalar $t$.  The relation
between $t$ and $\tau$ is 
\begin{equation}
\tau  = i + i t + {\cal O}(t^2) \,.
\label{eq:tau}
\end{equation}

Superamplitudes in this theory are classified according to a
maximally-helicity-violating (MHV) degree, $k =0, \ldots,n-4$ and the
numbers $n_+$ and $n_-$ of particles in the $\Phi^+$ and $\Phi^-$
multiplets~\cite{CarrascoAnomaly}. We shall denote the $n$-point
$\text{N}^k\text{MHV}^{(n_+,n_-)}$ amplitudes as:
\begin{equation} M^{(n_+,n_-)}_{n,k}\equiv 
                 M_{n,k}(\Phi_1^+, \ldots, \Phi_{n_+}^+, \Phi_{n_+ +1}^-,
  \ldots, \Phi_{n}^-)\,, 
\end{equation} 
where $n = n_+ + n_-$.
Only the ${\rm N}^k{\rm MHV}^{(n-k-2, k + 2)}$ tree-level scattering
superamplitudes of this theory are nonvanishing.  This is a consequence of the
$U(1)$ symmetry, which assigns charges $(0, \pm 1/2, \pm 1, \pm 3/2, -2, 2)$ to
the states $(h^{\pm\pm}$, $\!\psi^{\pm}$, $\!A^{\pm}$, $\chi^{\pm}$, $\!t$,
$\!\bar t)$ in \eqn{Multiplets}.
The $U(1)$ charges of spinors and vectors identify it as the on-shell form of
the third (anomalous) generator.
The main consequence of the anomaly in the S-matrix is that the selection rule
does not hold at loop level and all the amplitudes become
nonvanishing~\cite{CarrascoAnomaly}.

From the double-copy perspective, the two multiplets in \eqn{Multiplets}
correspond to the tensor products of an $\NeqFour$ super-Yang--Mills multiplet
and a positive or negative-helicity gluon state of a pure Yang--Mills theory.
In this way, the MHV degree $k$ corresponds to the one of the supersymmetric
side of the double copy and the $n_+$ and $n_-$ labels refer to the two
helicities of the gluon on the pure Yang--Mills side. The  ${U}(1)$ charge of a
given state is given by the difference of the helicities on the two sides of
the double copy: $q_{U(1)} =\text{ h({YM}) - h({SYM})}$.  The anomalous tree
amplitudes with $n_- = 0, 1, n-1$ or $n$ vanish trivially because the
corresponding pure Yang--Mills tree amplitudes vanish in the double copy. The
other cases are less trivial and rely on identities between gauge-theory
amplitudes, such as those described in Ref.~\cite{QuadraticRelations}.

\Section{Amplitudes}
One-loop scattering amplitudes in $\NeqFour$ supergravity have been
studied in Refs.~\cite{Dunbar1995,N4Oneloop, Dunbar2011dw, DunbarSoft, TourkineVanhove,
  CarrascoAnomaly}.  Following Ref.~\cite{N4Oneloop}, we
straightforwardly obtain all the anomalous four- and five-point
gravity amplitudes from the gauge-theory ones~\cite{BDKFourFivePoint}
using the double-copy construction~\cite{BCJ,BCJLoop}.  We present them here in the
spinor helicity conventions of Ref.~\cite{ManganoParke}. 
For an $n$-point amplitude we omit the  overall factor of $(\kappa/2)^{n}/(4\pi)^2$
where $\kappa$ is the gravitational coupling and the conserved supermomentum is
denoted by  $Q^A = \sum_{j=1}^n \lambda^{\alpha}_j \eta_{j}^A$.
At four-points the two independent anomalous superamplitudes are
\begin{align}
   {M}_{4,0}^{(0,4)} & =  i \, \deltaQ \,,
    \nonumber\\
   {M}_{4,0}^{(1,3)}  & =  -i \frac{\spb1.2\spa2.3\spa2.4}{\spa1.2\spa1.3\spa1.4} \, \deltaQ \,.
   \label{eq:4ptanom}
 \end{align}
Similarly, the five independent anomalous five-point superamplitudes are given by
\begin{align}
  {M}_{5,0}^{(0,5)} & = i \,2 \,\deltaQ \,,\nn \\
  {M}_{5,0}^{(1,4)} &= -i \sum_{r=2}^3 \frac{\spb1.r\spa r.4\spa r.5}{\spa1.r\spa1.4\spa1.5} \,\deltaQ \,, \nn\\
  {M}_{5,0}^{(2,3)} &= -i \varepsilon(1,2,3,4) \frac{\spa3.4^2\spa4.5^2\spa5.3^2}
                  {\prod_{i<j}\spa{i}.{j}} \, \deltaQ \,, \nn\\
  {M}_{5,0}^{(4,1)} & =  - i \frac{ s_{12}s_{34} \deltaQ}{\spa1.4\spa4.3\spa3.5\spa5.2 \spa2.1} 
        \biggl( \frac{2\spb1.4^2\spb1.3\spb2.4}{\spa2.3\spb4.5\spb1.5}\nn\\
 & \null 
  - \frac{\spb1.2^2\spa2.5^2\spa5.1^2 + \spb3.4^2\spa4.5^2\spa5.3^2}{\spa1.2\spa2.3\spa3.4\spa4.5\spa5.1} \biggr) \nn\\
& \hskip 2 cm    + ( 2 \leftrightarrow 3 ) \,, \nn\\
  {M}_{5,0}^{(5,0)} & = i \sum_{i<j} \frac{(\widehat{\gamma}_{ij})^2}{s_{ij}} \,\deltaQ \,,
  \label{eq:5ptanom05}
\end{align}
where $s_{ij}=(k_i+k_j)^2$, $\varepsilon(i,j,k,l) =
4i\varepsilon_{\mu\nu\sigma\rho} k_i^\mu k_j^\nu k_k^\sigma k_l^\rho$
and~\cite{FivePointN4}
\begin{equation}
  \widehat{\gamma}_{12}  = \frac{\spb1.2^2\spb3.4\spb4.5\spb3.5}{\varepsilon(1,2,3,4)}\,.
\end{equation}
The $(1,4)$, $(2,3)$ and $(4,1)$ amplitudes are new and the rest match
the results in Ref.~\cite{CarrascoAnomaly}. The result with $n_V$ vector multiplets running
in the loop is the same with an additional overall multiplicative factor of $(n_V+2)/2$.  As one would expect for
an anomaly in dimensional regularization, all of these amplitudes are
nonvanishing because of $\epsilon/\epsilon$ effects where $\epsilon =
(4-D)/2$.  It is also worth noting that through the lens of the
double-copy construction some of these amplitudes are obtained from
non-supersymmetric gauge-theory amplitudes which also have been
suggested to be nonzero because of another type of duality
anomaly~\cite{BardeenCangemi}.

In Ref.~\cite{CarrascoAnomaly}, the $n$-point amplitude,
\begin{equation}
  M_{n,0}^{(0,n)} = i (n-3)! \, \deltaQ \,,
  \label{eq:nptanom}
\end{equation}
is obtained via inverse-soft-scalar limits.  We have extended this to the
cases with $n_-\geq3$ and $n_+$ arbitrary, which correspond to all infinite
classes of MHV anomalous amplitudes except for those with $n_- = 0,1$.  All
superamplitudes of this kind are nonlocal, and can be obtained from the local
ones in \eqn{eq:nptanom} via the inverse-soft construction~\cite{InverseSoft}
which we implement using the soft-lifting functions of Ref.~\cite{DunbarSoft}.
A convenient way of presenting them is as the minors, with rows and columns
$M=\{m_1,\ldots,m_r\}$ removed~\cite{FengHeSoft},
$  S[M] = |\Phi|^{m_1\ldots m_r}_{m_1\ldots m_r}$,
of Hodges' $\Phi$ matrix~\cite{Hodges} with components
\begin{equation}
   \phi_i^j = \frac{\spb i.j}{\spa i.j}\;\;  \text{for } i\neq j\,, 
   \quad
    \phi_i^i = 
   -\sum_{j\neq i}\frac{\spb i.j\spa j.x \spa j.y }{\spa i.j \spa i.x \spa i.y } \ ,
\end{equation}
where $x,y$ are arbitrary reference spinors. We find that the $n_->3$
anomalous amplitudes are simply given in terms of 
the soft-lifting functions,
\begin{equation}
{M}_{n,0}^{(n_+,n_-)} = i (n_- -3)!\, S[M] \, \deltaQ\,,
  \label{eq:nonlocal}
\end{equation}
where $M$ is the set of $\Phi^-$ external states. The soft-lifting
function was used in Ref.~\cite{DunbarSoft} to obtain the rational
terms needed to complete the construction of the $n$-point
nonanomalous ($n_- = 2$) amplitudes.  We suspect that a similar
inverse-soft formula exists for the $n_- < 2$ amplitudes, but finding
it would require a detailed understanding of the kinematic
deformations that implement the inverse-soft procedure, which is
trivial for the $n_- \ge 3$ cases but not for the rest.  

\Eqn{eq:nonlocal} reproduces the corresponding results in
\eqns{eq:4ptanom}{eq:5ptanom05}. In addition, as explained in
Ref.~\cite{DunbarSoft}, the soft-lifting functions encode all the
correct soft and collinear limits involving the legs added in the
inverse-soft procedure. Indeed, one can straightforwardly check that
\eqn{eq:nonlocal} has the correct soft limits
\begin{align}
\begin{split}
  M_{n,0}^{(n_+,n_-)} & \xrightarrow{k_i\rightarrow 0} \mathcal{S}_{p_i} \, M_{n-1,0}^{(n_+-1,n_-)}\,,\qquad \quad \,\text{for}\quad \Phi_i^+\,, \\
  M_{n,0}^{(n_+,n_-)} & \xrightarrow{k_i\rightarrow 0} (n_--3)M_{n-1,0}^{(n_+,n_--1)}\,, \quad \text{for}\quad \Phi_i^-\,,
  \label{eq:softAmp}
\end{split}
\end{align}
where $\mathcal{S}_{p_i}=\phi_i^i$ is the usual graviton leading soft
factor.  Similarly, taking the supersymmetric collinear limits
\begin{align}
  \begin{split}
(\lambda_a,\tilde\lambda_a,\eta_a) &\rightarrow \sqrt{z}\,(\lambda_K,\tilde\lambda_K,\eta_K)\,, \\
(\lambda_b,\tilde\lambda_b,\eta_b) &\rightarrow \sqrt{1-z}\,(\lambda_K,\tilde\lambda_K,\eta_K)\,,
  \end{split}
\end{align}
we find that the amplitudes in \eqref{eq:nonlocal} have a universal phase singularity, i.e.,
\begin{align}
  \begin{split}
  M_{n,0}(\ldots,\, &\Phi_a^{h_a},\Phi_b^{h_b},\ldots) \xrightarrow{a || b} \\
  & \sum_{h_K=\pm} {\rm Sp}_{-h_K}^{h_a\,h_b}\,M_{n-1,0}(\ldots,\Phi_P^{h_K},\ldots) \,,
  \label{eq:collinear}
  \end{split}
\end{align}
where the $h_i$ denote the supermultiplet and the relevant splitting
functions are
\begin{equation}
  {\rm Sp}_{-}^{++} = - \frac{1}{z(1-z)} \frac{\spb a.b}{\spa a.b}\,, \qquad 
  {\rm Sp}_{+}^{-+} = - \frac{z}{(1-z)} \frac{\spb a.b}{\spa a.b} \,.
  \label{eq:splitting}
\end{equation}
As usual for gravity collinear limits with real momenta, we are only concerned
with the terms that contain phase singularities. (See Ref.~\cite{CollinearGravity} for 
further details.)  These checks fall short of a proof of \eqn{eq:nonlocal},
but as usual they give us confidence that this formula is correct.


\Section{Cancellation of anomalous amplitudes}
With these results in hand we can address the question of whether the anomalous
amplitudes can be cancelled by a local counterterm.
Ref.~\cite{CarrascoAnomaly} noted that the local amplitudes \eqref{eq:nptanom}
can be interpreted as arising from the following local $U(1)$-breaking terms in the
one-loop effective action,
\begin{align} 
\hskip -.1 cm 
\Gamma^\text{local}_{\rm  U(1)} & = \frac{1}{2(4\pi)^2}\int d^4x\, 
 \bigl(  (1-\log(1-\bar{t})) (R^+)^2\nn \\
& \hskip 1 cm \null 
+  (1-\log(1-t)) (R^-)^2 \bigr)\, + \,\hbox{SUSY}\,,
  \label{eq:countertermaction}
 \end{align}
where $R^+$ and $R^-$ are the self-dual and anti-self-dual parts of
the Riemann tensor, $ R^\pm_{\mu\nu\rho\sigma}= \pm\frac{i}{2}
\varepsilon_{\mu\nu}{}^{\alpha\beta} R^\pm_{\alpha\beta\sigma\rho}$
with $\varepsilon_{0123} = +1$. Again, in the presence of $n_V$ vector multiplets there is an extra
factor of $(n_V+2)/2$ in the coefficient.


As usual, definitions of quantum theories are ambiguous up to the
addition of local counterterms.  Ambiguities can be fixed by
demanding that classical symmetries and associated constraints on the
scattering amplitudes are preserved.  In this spirit, we choose to
define ${\cal N}=4$ supergravity to include the finite
local counterterm $S_{\rm ct} = -\Gamma^\text{local}_{U(1)}$ 
which subtracts away the local part of the anomalous effective action \eqref{eq:countertermaction}.
While this subtraction effectively changes the gauge of the auxiliary
$U(1)$ symmetry and thus departs~\cite{CarrascoAnomaly} from the
original formulation of the off-shell theory~\cite{N4superGrav}, 
it also sets to zero all
one-loop local anomalous amplitudes in~\eqn{eq:nptanom}, so 
it is the one we desire.

Deformations of the classical action by a local operator also
contribute to nonlocal amplitudes.  Since the counterterm sets to zero
the anomalous local amplitudes, it is interesting to consider its
effect on the nonlocal ones as well.
The presentation of the $n_- > 3$ amplitudes in \eqn{eq:nonlocal} makes it
clear that the same counterterm in \eqn{eq:countertermaction} also cancels this entire
class of amplitudes, because it cancels the seed $M_{n,0}^{(0, n)} $ of the
inverse soft construction.  
That is, for this class of superamplitudes, at one loop we have
\begin{equation}
M_{n,0}^{(n_+, n_-), \text{full}} =  M_{n,0}^{(n_+, n_-)} + M^{(n_+, n_-), S_\text{ct} }_{n, 0} = 0 \ ,
\label{AnomalyCancel}
\end{equation}
where  $M^{(n_+, n_-), S_\text{ct} }_{n, 0} $ are tree superamplitudes with a single vertex from the counterterm action 
in \eqn{eq:countertermaction}.

To independently confirm this cancellation, and to also check the fate of the
two amplitudes in \eqn{eq:5ptanom05} which are not given by \eqn{eq:nonlocal},
we compute the contribution of the counterterm action to one-loop amplitudes 
using the double-copy construction applied to higher-dimension operators~\cite{BroedelDixon}. 
Refs.~\cite{BroedelDixon, HeZhangHigherDim} show that,  when applied to 
two copies of pure Yang--Mills theory with a $\textrm{Tr}(F^{\mu}{}_\nu
F^{\nu}{}_\rho F^\rho{}_\mu)/3$ deformation, the double-copy approach yields the amplitudes in 
gravity deformed by $\phi^m R^2$ operators, where $\phi$ is a scalar field.
Using instead ${\cal N}=4$ super-Yang--Mills theory as the first
factor leads therefore to the tree-level amplitudes of ${\cal N}=4$
supergravity deformed by an operator $\mathcal{O}$ which is a
  supersymmetric completion of the $\phi^m R^2$ operators.
Using the $F^3$-deformed Yang--Mills amplitudes found in 
Refs.~\cite{DixonShadmiHigherDimension, BroedelDixon}, the 
Kawai--Lewellen--Tye (KLT)~\cite{KLT} double-copy formulae give
\begin{equation}
  \quad M^{(0,n),\mathcal{O}}_{n,0} = i (n-2)! \, \deltaQ \,,
  \label{eq:nptKLT}
\end{equation}
for $n=3,4,5$. Thus, at least for this number of external lines, the double-copy
yields the matrix elements of the supersymmetrization of the operator
\begin{equation}
  \mathcal{O} = \frac{1}{2 (4 \pi)^2} \Bigl((R^+)^2 \sum_n \bar t^n + (R^-)^2 \sum_n t^n \Bigr)\,,
\label{eq:cttermKLT}
\end{equation}
which only differs by a numerical factor from the expected counterterm
action \eqref{eq:countertermaction}: the number of
scalars. The normalization of these matrix elements can be changed to obtain those of \eqref{eq:countertermaction}, thus confirming through an
independent calculation that all three-, four- and five-point
anomalous one-loop amplitudes are cancelled after deforming the classical action
by the counterterm~\eqref{eq:countertermaction}.
Once lower-point amplitudes are cancelled by a counterterm, consistency
with soft and collinear graviton limits suggests that the cancellation
continues to all multiplicities, even for $n_- = 0,1$.
These results strongly indicate that the counterterm action cancels
both the local and the nonlocal one-loop anomalous amplitudes and thus
restores the $U(1)$ symmetry.  We expect that the addition of this
counterterm moves the anomaly into other generators of $SU(1,1)$
which do not appear to impose any selection rule on the scattering
amplitudes.

In addition, Ref.~\cite{CurvatureSquareN4} explains that the nonanomalous 
four-point superamplitude has the form
\begin{equation}
M^{(2,2)}_{4,0} = M^{\rm tree}_{R^2} + \cdots \ ,
\end{equation}
where $M^{\rm tree}_{R^2}$ are evanescent matrix elements of the
Gauss--Bonnet operator and its supersymmetric
completion~\cite{SChernSimons}. The same paper points out that the
anomalous and evanescent contributions are intertwined by the double
copy.  The local term in \eqn{eq:countertermaction} reflects this
observation.  Indeed, the $t$-independent term in \eqn{eq:countertermaction}
is the Gauss--Bonnet combination in the effective action,
\begin{equation}
\Gamma^\text{local}_{U(1)} = \frac{1}{2(4\pi)^2}\int d^4x\, R^* R^*  + \cdots\,.
\label{GaussBonnet}
\end{equation}
Counterterm \eqref{eq:countertermaction} removes both
the anomalous amplitudes and the evanescent contribution found at one
loop in Ref.~\cite{CurvatureSquareN4}.

Ref.~\cite{HuangSoft} noted that the non-anomalous amplitudes can give
rise to anomalous ones in the soft-scalar limit. This fact was
interpreted as a consequence of the duality anomaly. Following the analysis
above, we checked that the only effect of the counterterm on the four- and
five-point amplitudes is restricted to the cancellation of evanescent
pieces.  In particular, one can check that the soft limits are
unmodified.  In general higher-derivative corrections explicitly break the
noncompact duality symmetry resulting in nonvanishing soft
limits~\cite{ElvangKiermaier}.  In light of this, the connection
between soft scalar limits and the anomaly in the presence of counterterms
requires further analysis.

\Section{Conclusions}
In summary, in this letter we showed that both the local and nonlocal
anomalous MHV one-loop amplitudes of $\NeqFour$ supergravity can be
systematically cancelled by adding a local counterterm to the
classical action. These cancellations are nontrivial and strongly
suggest that all anomalous amplitudes in the theory can be removed 
in the same way.
Nevertheless, a number of issues remain.  Firstly, an off-shell
supersymmetric completion of our counterterm might help shed light on
the cancellations beyond scattering amplitudes.  A related
supersymmetric counterterm has been described in Ref.~\cite{Bernard};
it would be worthwhile to directly compare the anomalous matrix
elements of $\NeqFour$ supergravity with ones generated by this term,
as well as the one described in Ref.~\cite{BHSAnomalies}.
It would also be interesting to understand the relation, if any,
to other instances of cancellations of similar anomalies in other
four-dimensional field and string-theory models 
\cite{DTerms,GreenSchwarzSUGRAGravGauge,AnomStringN1,AnomStringN4}.
Perhaps more importantly, we would like to investigate the anomalous
amplitudes at higher loops.  Ref.~\cite{FourLoopN4Sugra} found that,
at four loops, both anomalous and nonanomalous amplitudes in
$\NeqFour$ supergravity carry a leading ultraviolet divergence.
This is surprising because the integrands of the anomalous amplitudes
must vanish in strictly four dimensions and therefore carry an extra
factor the dimensional regularization parameter $\epsilon$ compared to
the nonanomalous ones.  One would then expect the divergences of the
anomalous amplitudes to be suppressed, unless the actual source of the
divergence is the anomaly.
As mentioned above, the anomalous lower-loop amplitudes are expected to induce divergences at higher loops, 
even in the nonanomalous amplitudes~\cite{CarrascoAnomaly}.   Thus, the local counterterms removing the 
former should contribute nontrivially to the divergences of the latter.
The complete four-loop divergence should thus be reanalyzed.
To determine the counterterm effects on the four-loop divergence it is
necessary to evaluate anomalous amplitudes at higher loops and
understand whether further finite counterterms are necessary for their
removal. In any case, there are clearly new lessons to be
learned by investigating the higher-loop amplitudes of this theory.

\vskip .3 cm

\noindent {\it Acknowledgments:}
We thank Wei-Ming Chen, Clifford Cheung, Jared Claypoole, Lance Dixon, Thomas Dumitrescu, Yu-tin
Huang, Henrik Johansson, David Kosower, Oliver Schlotterer, Chia-Hsien Shen, Arkady Tseytlin,
Congkao Wen and Edward Witten for many enlightening discussions. This work was
supported in part by the Department of Energy under Award Numbers DE-SC0009937,
DE-SC0013699 and in part by the National Science Foundation under Grant Number NSF
PHY-1125915. We thank the Kavli Intitute for Theoretical Physics, where this
work was initiated, for hospitality.  J.P.-M. is supported by the U.S.
Department of State through a Fulbright Scholarship.  J.P.-M. also thanks the
Mani L. Bhaumik Institute for generous summer support.

\end{document}